\documentclass[twocolumn]{jpsj3}

\usepackage{graphicx}
\usepackage{mathptmx}    
\usepackage{color}
\usepackage{float}
\usepackage{amsbsy}
\usepackage{amsfonts}
\usepackage{amsmath}
\usepackage{amssymb}
\usepackage{latexsym}
\usepackage{mathrsfs}
\usepackage{bbm}
\usepackage{bm}
\usepackage[T1]{fontenc}
\usepackage{lmodern}

\newcommand{\ve}[1]{ \boldsymbol #1}
\newcommand{\bra}[1]{\langle #1 |}
\newcommand{\ket}[1]{| #1 \rangle}

\title{Outcome Independence of Entanglement in One-Way Computation}

\author{\name{Toshihiko \surname{Sasaki}}\thanks{Email: tsasaki@post.kek.jp. Present address: Photon Science Center of the University of Tokyo, 7-3-1 Hongo, Bunkyo-ku, Tokyo 113-8656, Japan.}, and \name{Tsubasa \surname{Ichikawa}}$^{1}$\thanks{Present address: Department of Physics, Gakushuin University, 1-5-1 Mejiro, Toshima-ku, Tokyo 171-8588, Japan.}, and \name{Izumi \surname{Tsutsui}}$^{2}$}

\inst{ Department of Physics, University of Tokyo, 7-3-1 Hongo, Bunkyo-ku, Tokyo 113-0033, Japan\\
$^{1}$ Research Center for Quantum Computing, Interdisciplinary Graduate School of Science and Engineering, Kinki University, 3-4-1 Kowakae, Higashi-Osaka, Osaka 577-8502, Japan\\
$^{2}$ Theory Center, Institute of Particle and Nuclear Studies, High Energy Accelerator Research Organization (KEK), 1-1 Oho, Tsukuba, Ibaraki 305-0801, Japan
}

\abst{We show that, in the standard scheme of one-way computation where the quantum circuit
consists of controlled-NOT gates and rotation gates, all intermediate states appearing in the process at
a given step of measurement are equivalent modulo local unitary transformations.
This implies, in particular, that all those intermediate states share the same entanglement irrespective of the measurement outcomes, indicating that the process of one-way computation is essentially unique with respect to local quantum operations.
}

\kword{one-way quantum computation, cluster state, entanglement, quantum computer, quantam information}

\begin{document}
\maketitle

\section{Introduction}

Entanglement is a key 
ingredient to render the  \lq quantum\rq\ distinctive against the \lq classical\rq.
The superiority of quantum computation ({\it e.g.} speed-up) over the classical counterpart, for instance, rests on the exploitation of entanglement, and 
it is essential for us to figure out how to achieve it effectively.
For implementation of quantum computation \cite{NielsenChuang200009,NakaharaOhmi200803}, two schemes have been primarily investigated; one is computation by synthesis of quantum logic gates \cite{Deutsch1989,Barenco1995}, and the other is one-way computation by local measurements of quantum states \cite{Briegel2001,Raussendorf2001,Raussendorf2003,Hein2006,Gross2009,Bremner2009}. 
The significance of entanglement in the former has been studied  \cite{Jozsa2003,Braunstein2002,Rungta2009}, and it is confirmed that entanglement is indeed crucial to achieve the increases in
computational power that quantum computing makes possible.
Meanwhile, for the latter scheme it is found \cite{Gross2009,Bremner2009} that, among all entangled states, cluster states provide a preferable basis for the increases.

One-way computation has a notable affinity with entanglement in that it consumes entanglement in local measurements.
This prompts us to ask precisely how entanglement is created and consumed in the actual process of computation.
However, this question has been deemed difficult to answer, because the process involves various intermediate states (IMS) generated by local measurements.   In fact, since the number of different IMS grows exponentially as the number of measurements increases, one may expect that 
the analysis of entanglement is virtually impossible.  

In this article, we show that this is not the case -- specifically, we prove that for one-way computation realized by a standard quantum circuit 
consisting of controlled-NOT (CNOT) gates and rotation (ROT) gates,
all IMS with different measurement outcomes are related by {\it local unitary transformations}.  In this respect, we recall that the outputs of each gate, which are special cases of IMS, are known to be related by local unitary transformations called byproduct operators \cite{Raussendorf2001}.
However, when one wants to examine the outcomes of general intermediate measurements, performed not necessarily in the order of gates prescribed for the computation, or even for partial set of qubits in the gates,  
the byproduct operators are no longer sufficient to  guarantee the local unitary
equivalence.   Our result holds in this most general case as well.  
Since entanglement is invariant under such transformations, this implies that the consumption process of entanglement in one-way computation is actually {\it unique}, irrespective of the outcomes of the measurements.

\section{Preliminaries}

To recall the prerequisite of one-way computation, consider an $n$-qubit system whose constituent qubits are labeled by $V=\{1,2,\cdots,n\}$.   Elements of the set $V$ may be regarded as vertices on a plane, where edges are formed by connecting two pairs  $i,j\in V$ we choose.   A {\it graph} $G(V, E)$ is then defined as the union of $V$ and the set $E$ of edges chosen.
Each vertex $i$ in the graph $G$ has the neighbor
$
 N_i = \{j\in V\mid \{i,j\}\in E\}
$
connected by the edges.
We may divide $V$ into three mutually exclusive subsets $V = C_{I}\cup C_{M}\cup C_{O}$, where $C_{I}$, $C_{M}$, and $C_{O}$ are called \lq input\rq, \lq middle\rq, and \lq output\rq\ section, respectively, such that the number of the vertices in $C_I$ is equal to that of $C_O$.    Each qubit represented by the vertex $i$ carries the Hilbert space ${\cal H}_i=\mathbb{C}^2$, and accordingly any set of vertices has the corresponding space given by the tensor product of the constituent ${\cal H}_i$.  For example,  the input section $C_I$ has ${\cal H}(C_I)=\bigotimes_{i\in C_I}{\cal H}_i$, and as a space it is identical to the  logical qubit space ${\cal H}(C_I) = {\cal H}_{\rm log}$ in which a desired unitary gate $U_{\rm desired}$ is realized.   The basic idea of one-way computation is to acquire the output state $U_{\rm desired} \ket{\psi_{\rm in}}$ in $C_O$ to a given input state $\ket{\psi_{\rm in}}$ in $C_I$, thereby achieving   
$\ket{\psi_{\rm in}} \to U_{\rm desired} \ket{\psi_{\rm in}}$ in ${\cal H}_{\rm log}$.  

For the actual implementation, we first prepare each of the qubits $i$ not belonging to $C_{I}$ ({\it i.e.}, $i \in V\backslash C_{I}$) in the $+1$ eigenstate $\ket{+}_i$ of  the spin operator $\sigma_x^i$ in ${\cal H}_i$.  Thus our  initial $n$-qubit state is
\begin{equation}
 \label{instate}
 \ket{\Psi_0}= \ket{\psi_{\rm in}}\otimes \bigotimes_{i\in V\backslash C_{I}} \ket{+}_i.
\end{equation}
Let $\mathbbm{1}^i$ be the identity operator on ${\cal H}_i$, and $\ket{0}_i$, $\ket{1}_i$ be the $+1$, $-1$ eigenstates of $\sigma_z^i$, respectively.    The conditional phase gate associated with the edge $\{i,j\}\in E$ reads
\begin{equation}
 \label{interactionS1}
  S_{ij} =\ket{0}_{ii\!}\bra{0}\otimes\mathbbm{1}^j+\ket{1}_{ii\!}\bra{1}\otimes \sigma_z^j.
\end{equation}
The graph state $\ket{G}$ corresponding to $G(V,E)$
is defined from the initial state by applying the conditional phase gate for all edges in the graph:
\begin{eqnarray}
 \label{clstate}
\ket{G} = S \ket{\Psi_0}, \qquad S = \prod_{\{i,j\}\in E} S_{ij}.
\end{eqnarray}
For brevity we hereafter omit the symbols $\otimes$ and $\mathbbm{1}^i$ when no confusion arises.
Note that $S$ satisfies
\begin{equation}
 \label{exrel}
  K_iS=S\sigma_x^i,
\qquad
K_i =\sigma_x^i\bigotimes_{j\in N_i}\sigma_z^j.
\end{equation}
It then follows from (\ref{instate}), (\ref{clstate}), and (\ref{exrel}) that
\begin{equation}
\label{eigeneq}
 K_i\ket{G}=\ket{G},
\end{equation}
for all $i\in V\backslash C_I$ \cite{Briegel2001,Raussendorf2001,Raussendorf2003,Hein2006}. 

Suppose that as an intermediate step of the one-way computation, we measure the spin of the $i$-th qubit in the $x$-$y$ plane with angle $\theta$ using the operator $\sigma_x^i \cos\theta + \sigma_y^i\sin\theta$.
According to the measurement outcomes $s=\pm1$, the state undergoes the change 
$
\ket{G} \to P_s^i(\theta) \ket{G}, 
$
where the acquired IMS is characterized by the projector,
\begin{eqnarray}
P_s^i(\theta) =\frac{\mathbbm{1}^i+s\left(\sigma^i_x\cos\theta+\sigma^i_y\sin\theta\right)}{2},
\end{eqnarray}
which fulfills
\begin{equation}
 \label{commutationThetaX1}
 P_s^i(\theta)\sigma_x^i = \sigma_x^i P_s^i(-\theta),
  \qquad
 P_s^i(\theta)\sigma_z^i = \sigma_z^i P_{-s}^i(\theta).
\end{equation}
From these we have
\begin{eqnarray}
P_s^i(\theta)K_j=
\begin{cases}
K_iP_s^i(-\theta)      &  \text{if $i=j$},\\
K_jP_{-s}^i(\theta)      & \text{if $i\neq j$, $i\in N_j$}, \\
K_jP_s^i(\theta)      & \text{if $i\neq j$, $i\not\in N_j$}.
\end{cases}
\label{PU}
\end{eqnarray}

\section{Local Unitary Equivalence}

Since our one-way computation consists of a set of ROT gates and CNOT gates, we first argue that these two admit independently
the local unitary equivalence for IMS, before combining the results to 
show that the same is true for a generic one-way computation.   

\subsection{Rotation Gate}

\begin{figure}[t]
 \begin{center}
  \includegraphics[width=2.8in]{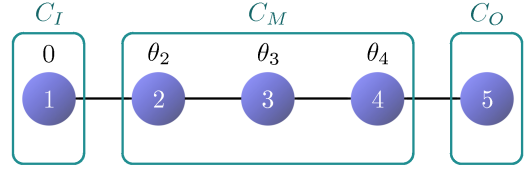}
 \end{center}
 \caption{(Color online) The graph $G_{\rm ROT}$ for the ROT gate.   The lines between the numbered vertices represent the edges, and we have, {\it e.g.}, the neighbor $N_2 = \{1, 3\}$.
 Measurement angles $\theta_i$ above the vertices $i$ are specified by Eq.~(\ref{order}).}
 \label{ROTgraph}
\end{figure}

Let us start with the one-qubit ROT gate, which can be parameterized by the Euler angles $\ve{\xi}=(\xi,\eta,\zeta)$ as
\begin{equation}
 U_{\text{ROT}}(\ve{\xi}) =
	\exp\left[-i\zeta\frac{\sigma_x}{2}\right]
	\exp\left[-i\eta\frac{\sigma_z}{2}\right]
	\exp\left[-i\xi\frac{\sigma_x}{2}\right].
	\label{urot}
\end{equation}
This gate can be implemented by the $n=5$ cluster state with the graph $G_{\rm ROT}$ shown in Fig.\ref{ROTgraph}.  
Let $s_i=\pm1$ be the outcomes of measurement for the $i$-th qubit with angle $\theta_i$, which is performed successively by the ascending order of $i$.
The actual measurement axis $\theta_i=\theta_i(\ve{\xi}, \ve{s})$ 
is determined from
the Euler angles $\ve{\xi}$ in the ROT gate and the measurement outcomes $\ve{s}=\{s_1, s_2, s_3\}$ as
 \begin{eqnarray}
\!\!\!\!\!
\theta_1=0,
\quad
\theta_2=-s_1\xi,
\quad
\theta_3=-s_2\eta,
\quad
\theta_4=-s_1s_3\zeta.
\label{order}
\end{eqnarray}

The measurement of the 1st qubit on the graph state (\ref{clstate}) yields the IMS $P_{s_1}^1(0)\ket{G_{\rm ROT}}$, which fulfills
\begin{equation}
P_{s_1}^1(0)\ket{G_{\rm ROT}}=P_{s_1}^1(0)K_2\ket{G_{\rm ROT}}=K_2P_{-s_1}^1(0)\ket{G_{\rm ROT}}, 
\label{ROTone}
\end{equation}
on account of Eqs.~(\ref{eigeneq}) and (\ref{PU}) with $1\in N_2$.  This
shows that the local unitary operator $K_2$ transforms an IMS to another IMS having the opposite measurement outcome.  
We also observe, from Eqs.~(\ref{eigeneq}) and (\ref{PU}) with $1\not\in N_3$ and $2\in N_3$, that the IMS obtained after the 2nd measurement obeys
\begin{eqnarray}
\begin{aligned}
P_{s_2}^2(\theta_2)P_{s_1}^1(0)\ket{G_{\rm ROT}}&=P_{s_2}^2(\theta_2)P_{s_1}^1(0)K_3\ket{G_{\rm ROT}}\\
&=P_{s_2}^2(\theta_2)K_3P_{s_1}^1(0)\ket{G_{\rm ROT}}\\
&=K_3P_{-s_2}^2(\theta_2)P_{s_1}^1(0)\ket{G_{\rm ROT}}.
\end{aligned}
\label{ROTtwo}
\end{eqnarray}
A similar argument using $K_2$, instead of $K_3$ above, yields
\begin{equation}
 P_{s_2}^2(\theta_2)P_{s_1}^1(0)\ket{G_{\rm ROT}}=K_2P_{s_2}^2(-\theta_2)P_{-s_1}^1(0)\ket{G_{\rm ROT}}.
\label{ROTtwo-esc}
\end{equation}
Since $-\theta_2=-(-s_1)\xi$, we conclude from (\ref{ROTtwo}) and (\ref{ROTtwo-esc}) that IMS in the 2nd measurement with different outcomes can be related by combining $\{K_2, K_3\}$.

Generalizing our reasoning, we see that the IMS of the 3rd measurement with the outcome $(s_1,s_2,s_3)$ can also be transformed into any IMS with a different outcome  $(s_1^\prime, s_2^\prime, s_3^\prime)$ by an appropriate combination of local unitary transformations $\{K_2, K_3, K_4\}$.  Clearly, the number of choices of $K_i$ is $2^3$ which is just the number of all possible different outcomes.   An analogous result holds for the IMS in the 4th measurement with $(s_1, s_2, s_3, s_4)$.
To summarize, we find that for the ROT gate all the IMS appearing at any stage of the measurement can be transformed into each other by local unitary transformations.

\subsection{CNOT Gate}

\begin{figure}[t]
 \begin{center}
  \includegraphics[width=2.8in]{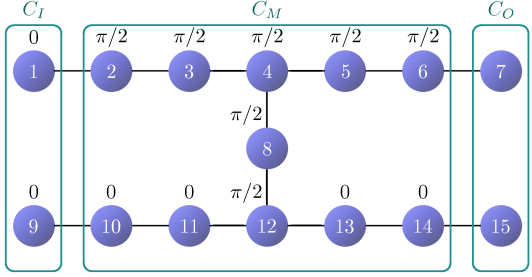}
  \end{center}
 \caption{(Color online) The graph $G_{\rm CNOT}$ for the CNOT gate.   Above the vertices $i$ the measurement angles $\theta_i$, which are either $0$ or $\pi/2$, are indicated. 
 }
 \label{CNOTgraph}
\end{figure}

Next we turn to the CNOT gate.  If implemented with $i$-th qubit as the control qubit and $j$-th as the target, the gate is represented by
\begin{equation}
U_{\text{CNOT}} =\ket{0}_{ii\!}\bra{0}\otimes\mathbbm{1}^j+\ket{1}_{ii\!}\bra{1}\otimes \sigma_x^j.
 \label{logicalCN}
\end{equation}
The gate, with the choice $i = 7$, $j = 15$, is realized by the $n=15$ graph $G_{\rm CNOT}$ shown in Fig.\ref{CNOTgraph}.   Unlike the ROT case (\ref{order}), all the measurement angles are predetermined as $\theta_i=0, \pi/2$ independently of the outcomes $s_j$.

Consider the local measurements over all qubits in $V\backslash C_O =C_I\cup C_M$.  
The IMS with the measurement outcomes $s_i$ are then given by
$
\prod_{i\in V\backslash C_O}P^{i}_{s_{i}}(\theta_{i})\ket{G_{\rm CNOT}}
\label{PS}
$
up to a normalization factor.
Using the identity
$
P_s^i(\theta)=P_{-s}^i(\theta+\pi)
$
and (\ref{PU}), we obtain
\begin{eqnarray}
K_i \prod_{j\in\{i\}\cup N_i}P^j_{s_j}(\theta_j) =
 P_{s_i}^i(0)\prod_{j\in N_i}P_{-s_j}^j(\theta_j) \, K_i
 \label{UPzero}
\end{eqnarray}
for $i$ with $\theta_i=0$, and
\begin{eqnarray}
K_i \prod_{j\in\{i\}\cup N_i}P^j_{s_j}(\theta_j)=
 \prod_{j\in\{i\}\cup N_i}P_{-s_j}^j(\theta_j) \, K_i
\end{eqnarray}
for $i$ with $\theta_i=\pi/2$. These relations show that the action of $K_i$ on the IMS flips the measurement outcomes on the qubits in $N_i$  (including $i$-th qubit for $\theta_i=\pi/2$) in IMS. In 
Table I, we summarize the sets of $s_j$ whose elements are flipped by $K_i$.  By combining these $K_i$ appropriately, we can construct 
unitary operators which flip the outcomes of a specific qubit without flipping the rest, 
which implies that 
all IMS can be related to each other by local unitary transformations.
The above argument also ensures that, by an appropriate local unitary operation, we can change the measurement outcomes freely even when not all of the qubits are measured.

\subsection{Universal Gate Set}

\begin{figure*}[t]
 \begin{center}
  \includegraphics[width=2.5in]{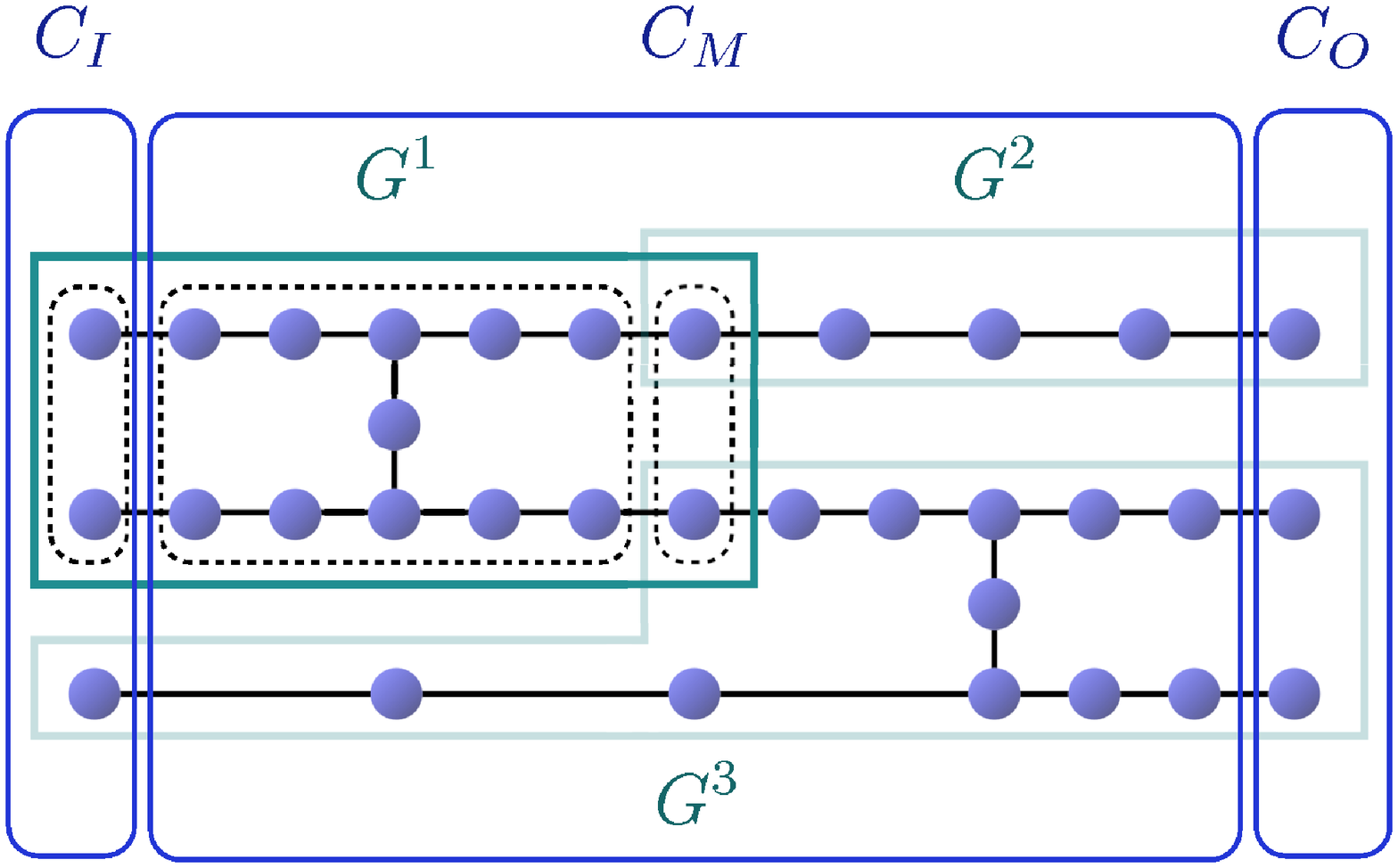}
  \qquad \quad
    \includegraphics[width=3.5in]{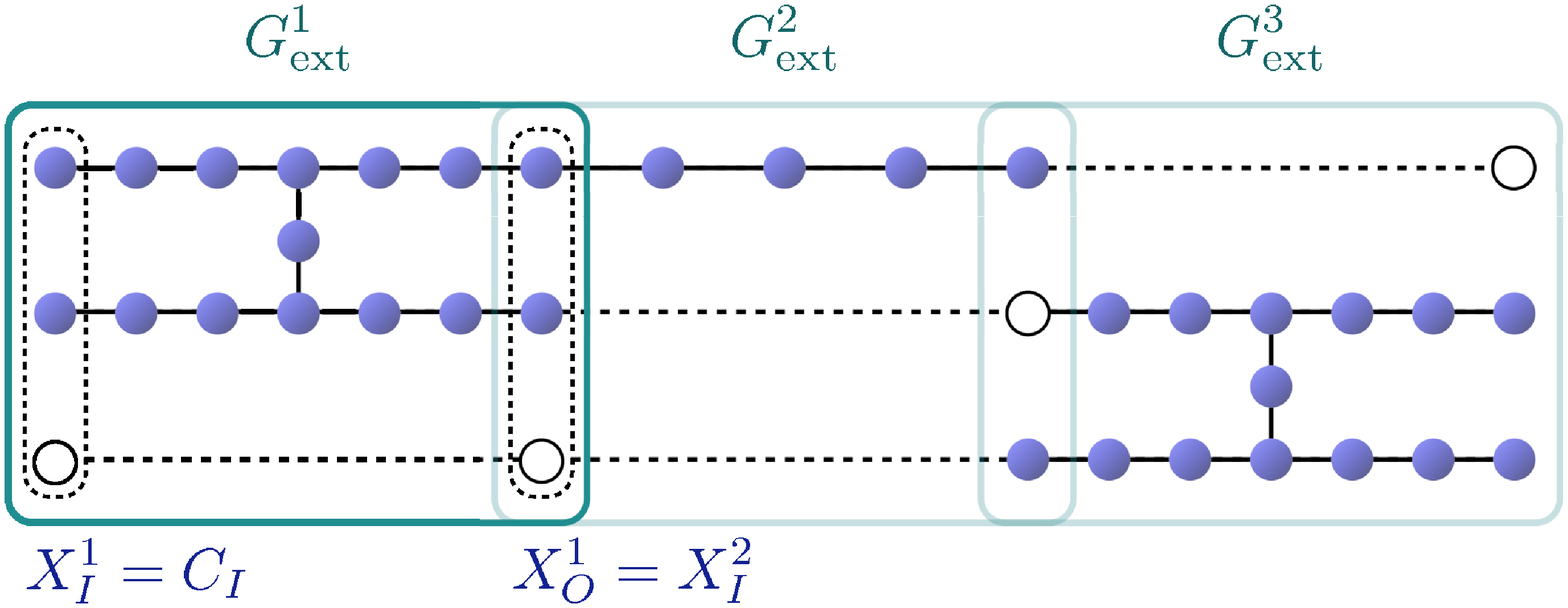}
  \end{center}
  \vskip -2.0ex
 \caption{ (Color online) (Above) The graph $G$ for the unitary gate $U_{\rm desired}=U_{\text{CNOT}} U_{\text{ROT}} U_{\text{CNOT}}$.    (Below)  The diagram of the process of computation with extended graphs $G^\alpha_{\rm ext}$ obtained by adding to $G^\alpha$ virtual vertices (open circles) which are aliases of the nearest vertices connected by the dotted lines.   All the input and output sections $X^\alpha_I$ and $X^\alpha_O$ in $G^\alpha_{\rm ext}$ possess the same number of qubits to provide the space ${\cal H}_{\rm log}$.
}
 \label{f3}
\end{figure*}

Now we come to the point to show that one-way computation for the universal gate set enjoys the same unitary equivalence.  To this end, recall first that in the logical space ${\cal H}_{\rm log}$ any unitary gate $U_{\rm desired}$ can be decomposed into a product of ROT and CNOT gates, 
\begin{equation}
U_{\rm desired}=U_{m}(\ve{\xi}^m)\, U_{m-1}(\ve{\xi}^{m-1})\, \cdots \,  U_{1}(\ve{\xi}^1),
\label{udec}
\end{equation}
where $U_\alpha(\ve{\xi}^\alpha)$, $\alpha = 1, \ldots, m$, are either $U_{\text{ROT}}$ in (\ref{urot}) or $U_{\text{CNOT}}$ in (\ref{logicalCN}) acting in (generally different) subspaces in ${\cal H}_{\rm log}$,  with $\ve{\xi}^\alpha=(\xi^\alpha, \eta^\alpha, \zeta^\alpha)$ being relevant only for $U_{\text{ROT}}$. 
Each $U_\alpha$ is implemented at step $\alpha$ in the whole process of computation and, accordingly, 
we consider a graph $G$ consisting of subgraphs $G^\alpha$, with their own vertices $V^\alpha =C_I^\alpha\cup C_M^\alpha\cup C_O^\alpha$, which are either $G_{\rm ROT}$ or $G_{\rm CNOT}$ in correspondence with $U_\alpha$ in (\ref{udec}).   The actual process of
step $\alpha$ involves an extended graph $G^\alpha_{\rm ext} \supset G^\alpha$ with extra vertices which are irrelevant for the implementation of $U_\alpha$ but necessary to provide ${\cal H}_{\rm log}$ as the operational space.   
We denote by $X_I^\alpha$ and $X_O^\alpha$  the input and the output section of $G^\alpha_{\rm ext} $ containing $C_I^\alpha$ and $C_O^\alpha$, respectively, for which we have ${\cal H}(X_I^\alpha)={\cal H}(X_O^\alpha)={\cal H}_{\rm log}$.
The input section $X_I^\alpha$ contains those vertices in $C_O^\beta$ with $\beta \le \alpha$ which have not been used in earlier steps, and likewise 
$X_O^\alpha$ contains those vertices in $C_I^\beta$ with $\beta \ge \alpha$ which will be used in later steps, such that $X_I^1 = C_I$, 
$X_O^k = X_I^{k+1}$ for $k = 1, \ldots, m-1$ and $X_O^m = C_O$ (see Fig.\ref{f3} for illustration).

\begin{table}[t]
  \begin{center}
      \caption{(Left)  The action $K_i$ and the flipped qubits $j$ in the measurement outcomes $s_j$.  (Right)  The qubit $i$ and the combined operator required to flip only the outcome $s_i$ leaving all the rest $s_j$ for $j \ne i$ unaltered. }
   \label{CN}
  \begin{tabular}{cl|cl}
\hline
    operator & flipped qubits$\,$  &  $\,$ qubit $\,$ & combined operator\\
\hline
\hline
  $K_2$   & $1, 2, 3$ &  $1$   & $K_2K_3K_5K_6$ \\
  $K_3$   & $2, 3, 4$ &  $2$   & $K_3K_4K_5K_7K_8K_{13}K_{15}$ \\
  $K_4$   & $3, 4, 5, 8$ &  $3$   & $K_4K_6K_7K_8K_{13}K_{15}$\\
  $K_5$   &  $4, 5, 6$ &  $4$   &  $K_5K_6$\\
  $K_6$   & $5, 6$ &  $5$   & $K_6K_7$\\
  $K_7$   & $6$ &  $6$   & $K_7$\\
$K_8$   & $4, 8, 12$ &$8$   & $K_5K_6K_8K_{13}K_{15}$\\
  $K_{10}$   & $9, 11$ &  $9$   & $K_5K_6K_8K_{10}K_{12}K_{14}$\\
  $K_{11}$   & $10,12$ &  $10$   & $K_{11}K_{13}K_{15}$\\
  $K_{12}$   & $8, 11, 12, 13$ &  $11$   & $K_5K_6K_8K_{12}K_{14}$\\
  $K_{13}$   & $12, 14$ &  $12$   & $K_{13}K_{15}$\\
  $K_{14}$   & $13$ &  $13$   & $K_{14}$\\
  $K_{15}$   & $14$ &  $14$   & $K_{15}$\\
\hline
\end{tabular}
\end{center}
   \smallskip
\end{table}

To describe the process more explicitly, consider a projection associated with the measurements over  
an arbitrary subset $L^\alpha \subset V^\alpha\backslash C_O^\alpha$ of qubits in $G^\alpha$ with outcomes 
$\ve{s}^\alpha =\{s^\alpha_i=\pm1\ |\ i\in L^\alpha\}$,
\begin{equation}
 P(L^\alpha , \ve{\xi}^\alpha, \ve{s}^\alpha)=\prod_{i\in L^\alpha}P_{s_i^\alpha}^i(\theta_i^\alpha),
\end{equation}
where $\theta_i^\alpha$ are given by $\theta_i^\alpha=\theta_i(\ve{\xi}^\alpha, \ve{s}^\alpha)$ as in (\ref{order}) for $G^\alpha=G_{\rm ROT}$, while $\theta_i^\alpha=0$ or $\pi/2$ for $G^\alpha=G_{\rm CNOT}$ according to Fig.~\ref{CNOTgraph}.   With modified angles $f^{\alpha}\ve{\xi}^\alpha$ (to be discussed shortly) with $f^{1}\ve{\xi}^1 = \ve{\xi}^1$, 
the IMS of the entire system at an intermediate step $\alpha = k$ after the measurements over $\Lambda_k = \cup_{\alpha = 1}^k L^\alpha$  can then be written as 
\begin{equation}
\ket{\Psi(\Sigma_k; \Lambda_k)} :=
\left[ \prod_{\alpha=1}^{k} P(L^\alpha, f^{\alpha}\ve{\xi}^\alpha,\ve{s}^\alpha) S^\alpha \right]\ket{\Psi_0},
\label{prjst}
\end{equation}
where the product is in the descending order of $\alpha$ from the left.   In (\ref{prjst}),
$S^\alpha=\prod_{\{i,j\}\in E^\alpha}S_{ij}$ is the operator (\ref{interactionS1}) associated with the edges $E^\alpha$ in $G^\alpha$,  
$\ket{\Psi_0}$ is the initial state (\ref{instate}) for the total graph $G$, 
and we have introduced the notation $\Sigma_{k}=\{\ve{s}^1,\ldots, \ve{s}^{k}\}$ for the collection of the measurement outcomes up to step $k$.  
The choice of $\Lambda_k$ is not completely free because we cannot measure a qubit whose measurement angle is not determined by the measurement results of previously measured qubits.
This means that the IMS in (\ref{prjst}) are those (and actually the most general) states which
appear in an actual process of the one-way computation, where the measurement angles $f^\alpha\ve{\xi}^\alpha$ of qubits in $\Lambda_k$ are determined by earlier measurement results of qubits in $\Lambda_k$.

We now notice that, by using the $k=1$ IMS in (\ref{prjst}), the local unitary equivalence argued earlier for ROT and CNOT may be expressed concisely as
\begin{eqnarray}
\begin{aligned}
 \label{general-graph-base-step}
 \ket{\Psi(\Sigma_1; \Lambda_1)} = U(\Sigma_1,\Sigma_1^\prime)\, \ket{\Psi(\Sigma_1^\prime; \Lambda_1)},
 \end{aligned}
\end{eqnarray}
with a local unitary transformation $U(\Sigma_1,\Sigma_1^\prime)$. 
Indeed, this is so because $\ket{\Psi_0}$ in (\ref{prjst}) contains $\bigotimes_{i\in \left(C_M^1\cup C_O^1\right)} \ket{+}_{i}$ which is sufficient for our argument there.

An important property in one-way computation is that,  after the full measurements $L^\alpha=V^\alpha\backslash C_O^\alpha$, 
the IMS at each step $k$ admits the form,
\begin{eqnarray}
\ket{\Psi(\Sigma_k; \Lambda_k)} = \ket{\psi_{\rm out}^k}\otimes \ket{\phi^k},
\end{eqnarray}
where $\ket{\psi_{\rm out}^k} \in {\cal H}(X_O^k)$ is the output state, and $\ket{\phi^k} \in {\cal H}(V\backslash X_{O}^k)$.
The output state $\ket{\psi_{\rm out}^k}$, which becomes the input state $\ket{\psi_{\rm in}^{k+1}}$ in the next step, turns out to be
\begin{eqnarray}
\ket{\psi_{\rm out}^k} = {R_k}\, U_k(f^{k}\ve{\xi}^k) \ket{\psi_{\rm in}^{k}},
\label{rfactor}
\end{eqnarray}
with a qubit-wise local unitary (byproduct) operator ${R_k}=R_k(\ve{s}^k)$,  where $\ket{\psi_{\rm in}^1}$ is given by $\ket{\psi_{\rm in}}$ in (\ref{instate}).   The maps 
$f^\alpha$ are then determined \cite{Raussendorf2003} from the demand that at the final step $m$ we obtain
\begin{eqnarray}
\ket{\psi_{\rm out}^m} 
&=&  \left[ {R_m} U_m(f^{m}\ve{\xi}^m)\cdots{R_1} U_1(f^{1}\ve{\xi}^1)\right]\ket{\psi_{\rm in}^1} \nonumber\\
&=& T\, U_{\rm desired} \ket{\psi_{\rm in}^1},
\label{desreal}
\end{eqnarray}
with some local unitary gate $T$.   

Having given the relationship between adjacent steps, it is straightforward to extend the result (\ref{general-graph-base-step}) to the final step $k= m$  (for detail, see the Appendix):
\begin{eqnarray}
\begin{aligned}
\label{general-graph-main-statement}
 \ket{\Psi(\Sigma_m; \Lambda_m)} = U(\Sigma_m,\Sigma_m^\prime)\, \ket{\Psi(\Sigma_m^\prime; \Lambda_m)}.
 \end{aligned}
\end{eqnarray}
This shows that any two IMS with different outcomes $\Sigma_m$ and $\Sigma_m^\prime$, obtained under the measurements on the same but arbitrary set $\Lambda_m$ which corresponds to an intermediate step in an actual process of the one-way computation, are equal up to a unitary local transformation $U(\Sigma_m,\Sigma_m^\prime)$.  The equivalence of entanglement possessed by those intermediate IMS follows immediately from this, on account of the general requirement of the unitary equivalence for entanglement measures \cite{Horodecki2009}.

\begin{figure*}[t]
 \begin{center}
    \includegraphics[width=3.5in]{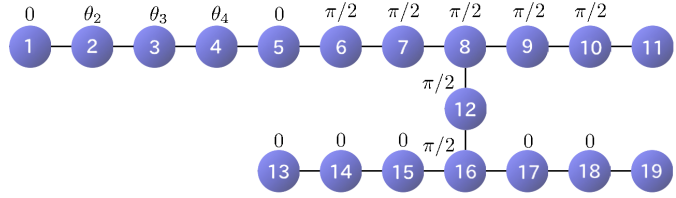}
  \end{center}
  \vskip -2.0ex
 \caption{ (Color online) The graph for the unitary gate $U_{\text{ROT}} U_{\text{CNOT}}$ possessing the input section $C_I=\{1,13\}$ and the output section $C_O=\{11,19\}$.   The measurement angles are indicated above the respective qubits.}
 \label{note-img}
\end{figure*}

\section{Summary and Discussions}

In this article, we have shown that, for the universal gate set consisting of ROT gates and CNOT gates, all IMS with different outcomes for an arbitrarily chosen set of measurements can be related by local unitary operations.
This rather simple observation should be 
handy for tracking the consumption process of entanglement in the cluster state during one-way computation.  For instance, this will reduce the complexity of evaluating multipartite entanglement measures ({\it e.g.}, those based on concurrence \cite{Ichikawa2009,Ichikawa2009a}) which are required to be invariant under local unitary transformations, allowing us to consider only a single IMS for each measurement.   Note that our equivalence is established for local unitary operations, not for LOCC (local operations and classical communication) under which entanglement measures are only monotone rather than invariant.
We hope that the essential uniqueness of IMS pointed out here provides a basis for comparing 
directly  the process of one-way computation with those of quantum logic gates, and thereby assists our understanding on quantum computation further.

\begin{acknowledgement}
TS is supported by JSPS Research Fellowships for Young Scientists, TI is supported by \lq Open Research Center\rq~Project for Private Universities: matching fund subsidy, and IT is supported by the Grant-in-Aid for Scientific Research (C), No.~20540391-H22, all of MEXT, Japan.  
\end{acknowledgement}

\section*{Note}

After this work was completed, we learned that a similar result was obtained in Refs.\cite{Danos2006,Browne2007}.  Their result
may be summarized as follows: if one assumes that 0) a \lq\lq Pauli flow\rq\rq~can be found on a graph and its
measurement pattern, then there exists a series of
measurements such that 
1) it can sends input pure states to output pure states, 
2) the measurement angles are determined by earlier measurement results,
and 
3) the outcome states with different measurement outcomes are equal to each other up to a local Clifford group.  

In their paper, the internal relations among  
the three assertions 1), 2) and 3) were not discussed and remained unclear, whereas 
in this article we have shown that 3) actually follows from 2) (which is used to ensure the existence of the function $f^\alpha$ in the Appendix) when
the measurement patterns are restricted to the circuits made of CNOT and ROT gates.
The difference in logical structure between the two is significant, since
assumption 0) is not trivial at all.  
 
To see this more explicitly, let us consider, {\it e.g.},  the unitary gate $U_{\text{ROT}} U_{\text{CNOT}}$ shown in Fig. \ref{note-img}.
After the measurements of five qubits numbered as 1, 2, 3, 5 and 6 (whose angles are determined by the original rules \cite{Raussendorf2003}),  we are left with fourteen-qubit entangled states as our IMS.
Depending on the measurement outcomes, there arise $2^5$ different types of entangled states.
Our result assures the local unitary equivalence among all these states, that is, assertion 3) from 2).
On the other hand, the result of Refs.\cite{Danos2006,Browne2007} assures the local unitary equivalence of such IMS only when there exists a corresponding Pauli flow with the partial order ``$<$'' satisfying $i\not < j$, $\forall i\in\{4, 7, 8, 9, 10, 11, 12, 13, 14, 15, 16, 17, 18, 19\}$, $\forall j\in \{1,2,3,5,6\}$. 
This is a highly nontrivial problem and does not seem to admit an immediate answer, even if it turns out to be affirmative.
In comparison,  our assumption 2) is rather mild from physical grounds and can always be checked by applying the rules \cite{Raussendorf2003} inductively.
Incidentally, our result may in fact suggest that 2) implies 0), which should also be interesting to confirm.

\appendix
\section{}
In this Appendix, we prove 
the local unitary equivalence (\ref{general-graph-main-statement}) of IMS by mathematical induction starting with (\ref{general-graph-base-step}).
Our argument will be similar to those given in the text, except for some technical complication due to 
the maps $f^\alpha$ which become nontrivial for $k  \ge 2$.   Prior to the proof, we describe $f^\alpha$ and also present two formulas to be used.

We put, for simplicity, all the unmeasured outcomes as $+1$, which is assumed to be possible here without influencing
the measurement outcomes over $\Lambda_k = \cup_{\alpha = 1}^k L^\alpha$.  
With $g_i = \frac{1-s_i}{2}$, 
the byproduct operators $R_{\alpha}(\ve{s}^\alpha)$ appearing in (\ref{rfactor}) under the given outcomes can be written as (see Ref. \cite{Raussendorf2003})
\begin{equation}
 \label{general-gate-def-sigma-rot}
 R_{\text{ROT}}=(\sigma_x)^{g_2+g_4}(\sigma_z)^{g_1+g_3},
\end{equation}
if $U_\alpha$ is ROT, and 
\begin{equation}
 \label{general-gate-def-sigma-cnot1}
 R_{\text{CNOT}} = (\sigma_x^{(c)})^{\gamma_x^{(c)}}(\sigma_x^{(t)})^{\gamma_x^{(t)}}(\sigma_z^{(c)})^{\gamma_z^{(c)}}(\sigma_z^{(t)})^{\gamma_z^{(t)}},
\end{equation}
if $U_\alpha$ is CNOT, where the  factors associated with the spin operators of the control and target qubits are given by
\begin{eqnarray}
 \gamma_x^{(c)} &=& g_2 + g_3 + g_5 + g_6, \nonumber\\
 \gamma_x^{(t)} &=& g_2 + g_3 + g_8 + g_{12}+ g_{14},\nonumber\\
 \gamma_z^{(c)} &=& g_1+g_3+g_4+g_5+g_8 +g_9 + g_{11}+1,\nonumber\\
 \gamma_z^{(t)} &=& g_9 + g_{11} + g_{13}.
\end{eqnarray}
We also record here some useful algebraic relations,
\begin{eqnarray}
 U_{\text{ROT}}[\xi,\eta,\zeta]\,\sigma_x &=&\sigma_x \,U_{\text{ROT}}[\xi,-\eta,\zeta], \nonumber\\
 U_{\text{ROT}}[\xi,\eta,\zeta]\,\sigma_z &=&\sigma_z \,U_{\text{ROT}}[-\xi,\eta,-\zeta], \nonumber\\
 U_{\text{CNOT}}\, \sigma_x^{(t)} &=& \sigma_x^{(t)} \,U_{\text{CNOT}}, \nonumber\\
 U_{\text{CNOT}} \,\sigma_x^{(c)} &=& \sigma_x^{(c)}\sigma_x^{(t)} \,U_{\text{CNOT}}, \nonumber\\
 U_{\text{CNOT}} \,\sigma_z^{(t)} &=& \sigma_z^{(c)}\sigma_z^{(t)} \,U_{\text{CNOT}}, \nonumber\\
 U_{\text{CNOT}}\, \sigma_z^{(c)} &=&\sigma_z^{(c)} \, U_{\text{CNOT}}.
\end{eqnarray}

Now, we set $T_1=\mathbbm{1}$ and define the gate $W_\alpha$ by
\begin{eqnarray}
W_\alpha=\begin{cases}
T_\alpha      & \text{if $U_\alpha$ is ROT}, \\
U_{\text{CNOT}}\, T_\alpha\, U_{\text{CNOT}}^{-1}      & \text{if $U_\alpha$ is CNOT},
\end{cases}
\end{eqnarray}
and then put $ T_{\alpha+1} = R_{\alpha}W_\alpha$ to proceed to the next step.  
This allows us to determine all these quantities for higher steps iteratively, and the maps $f^\alpha$ are defined by the relation,
\begin{equation}
 \label{appendix-commutation-simultaneous}
  U_\alpha(f^{\alpha}\ve{\xi}^\alpha)=W_\alpha U_\alpha(\ve{\xi}^\alpha)T_\alpha^{-1}.
\end{equation}
This in fact ensures (\ref{desreal}) with the unitary gate $T = T_{m+1}$.

At this point, we note that $T_\alpha$ is regarded as a local unitary operator on ${\cal H}(X_I^{\alpha}) (= {\cal H}_{\rm log})$, but it may be extended to a tensor product 
$\tilde{T}_\alpha:=O\otimes T_\alpha\otimes \mathbbm{1}$ acting on ${\cal H}(V)$, where $O$ is an element of the Pauli group on ${\cal H}(\bigcup_{i=1}^{\alpha-1}(C_I^i\cup C_M^i))$ and $\mathbbm{1}$ is the identity on the complementary subspace in ${\cal H}(V)$.  
The choice of $O$ is immaterial in our discussion, because it commutes with $P(X^\beta , \ve{\xi}^\beta, \ve{s}^\beta)$ and $ S^\beta$ for $\beta=\alpha,\cdots, m$.
Analogously, one can define $\tilde{W}_\alpha$ and $\tilde{R}_\alpha$  from $W_\alpha$ and $R_\alpha$ as the unitary operators on ${\cal H}(X_O^\alpha)$.

For these extended operators, we first show 
\begin{equation}
 \label{appendix-byproduct-relation1}
P(L^\alpha, f^{\alpha}\ve{\xi}^\alpha, \ve{s}^\alpha) S^\alpha \tilde{T}_\alpha \ket{\Psi_\alpha}
=\tilde{W}_\alpha P(L^\alpha , \ve{\xi}^\alpha, \ve{s}^\alpha) S^\alpha  \ket{\Psi_\alpha},
\end{equation}
for 
\begin{eqnarray}
\ket{\Psi_\alpha} = \ket{\phi_{\rm in}}\otimes\bigotimes_{i\in C_M^\alpha\cup C_O^\alpha}\ket{+}_i
\end{eqnarray}
with arbitrary
$
\ket{\phi_{\rm in}}\in{\cal H}(V\backslash (C_M^\alpha\cup C_O^\alpha)).
$
Indeed, if $U_\alpha$ is ROT, and if $T_\alpha = \sigma_z$, for example, then from (\ref{appendix-commutation-simultaneous}) we have $W_\alpha=T_\alpha$ and $f^\alpha\ve{\xi}^\alpha=(-\xi, \eta, -\zeta)$ for $\ve{\xi}^\alpha=(\xi, \eta,\zeta)$. 
Setting $\tilde{T}_\alpha=O\sigma_z^1$ and using (\ref{eigeneq}) and (\ref{commutationThetaX1}), we find
\begin{eqnarray}
 &&  P(L^\alpha,f^\alpha\ve{\xi}^\alpha,\ve{s}^\alpha)S^\alpha\tilde{T}_\alpha\ket{\Psi_\alpha}\nonumber\\
 &=& O P(L^\alpha,f^\alpha\ve{\xi}^\alpha,\ve{s}^{\alpha})\sigma_z^1 S^\alpha	
	\ket{\Psi_\alpha}\nonumber\\
 &=& O P(L^\alpha,f^\alpha\ve{\xi}^\alpha,\ve{s}^{\alpha})\sigma_z^1 K_2^\alpha K_4^\alpha S^\alpha	
	\ket{\Psi_\alpha}\nonumber\\
&=& O P(L^\alpha,(-\xi,\eta,-\zeta),\ve{s}^\alpha)\sigma_x^2\sigma_x^4\sigma_z^5 S^\alpha
	\ket{\Psi_\alpha}\nonumber\\
 &=& O\sigma_x^{2}\sigma_x^{4}\sigma_z^{5}
	P(L^\alpha,\ve{\xi}^\alpha,\ve{s}^\alpha)S^\alpha\ket{\Psi_\alpha},
\end{eqnarray}
where the numbers $\{1,2,3,4,5\}$ are the labels of qubits for ROT (see Fig.~\ref{ROTgraph}).
Since $C_O^\alpha=\{5\}$ for this case, we can put $\tilde{W}_\alpha=O\sigma_x^{2}\sigma_x^{4}\sigma_z^5$, which demonstrates (\ref{appendix-byproduct-relation1}).  Other choices of  $T_\alpha$ or the case of CNOT  can be discussed similarly.  

We also wish to establish
\begin{equation}
\label{appendix-result-of-subsectionAB}
 P(L^\alpha , \ve{\xi}^\alpha, \ve{s}^\alpha) S^\alpha \ket{\Psi_\alpha} 
=\tilde{R}_\alpha\tilde{R}'_\alpha P(L^\alpha , \ve{\xi}^\alpha, \ve{s}'^\alpha) S^\alpha \ket{\Psi_\alpha},
\end{equation}
as a generalization of (\ref{general-graph-base-step}).
Again, we examine this with an example, this time for $U_\alpha$ given by CNOT.  
Consider two sets of the measurement outcomes $\ve{s}$ and $\ve{s}^\prime$ with, say,  
$s_3\neq s'_3, s_i=s'_i (i\neq 3)$.  In this case, from (\ref{general-gate-def-sigma-cnot1}) we have $R_\alpha R'_\alpha = \sigma^{7}_x \sigma^{7}_z \sigma^{15}_x$, whereas from Table I, we find
\begin{eqnarray}
\!\!\!\!\!\!
U(\Sigma_\alpha, \Sigma^{\prime}_\alpha) = K_4K_6K_7K_8K_{13}K_{15}=O\sigma_{x}^{7}\sigma_z^7\sigma_x^{15}
\end{eqnarray} 
by choosing an appropriate operator $O$ in the Pauli group.  We thus find $U(\Sigma_\alpha, \Sigma^{\prime}_\alpha) = \tilde{R}_\alpha\tilde{R}'_\alpha$, which shows (\ref{appendix-result-of-subsectionAB}).  Other cases can also be argued analogously.

With these formulas (\ref{appendix-byproduct-relation1}) and (\ref{appendix-result-of-subsectionAB}), we
now prove (\ref{general-graph-main-statement}) for
\begin{eqnarray}
U(\Sigma_m, \Sigma^{\prime}_m) = \tilde{T}_{m+1}\tilde{T}'_{m+1},
\label{UTT}
\end{eqnarray}
based on the assumption,
\begin{eqnarray}
\ket{\Psi(\Sigma_\alpha; \Lambda_\alpha)}= \tilde{T}_{\alpha+1}\tilde{T}'_{\alpha+1}\ket{\Psi(\Sigma_\alpha^\prime; \Lambda_\alpha)}
\label{appendix-base-step}
\end{eqnarray}
for $\alpha = k-1$ with some $k$.  For $\alpha =1$ we have already this, because $T_2=R_1$ implies $\tilde{T}_2=\tilde{R}_1$ and hence (\ref{appendix-base-step}) with $\alpha =1$ follows from (\ref{general-graph-base-step}).   For $\alpha = k$, we utilize (\ref{appendix-byproduct-relation1}), (\ref{appendix-result-of-subsectionAB}) and ({\ref{appendix-base-step}) with $\alpha = k-1$ to observe

\begin{eqnarray}
  &&\ket{\Psi(\Sigma_k; \Lambda_k)}\nonumber\\
 &=& \left[P(X^k,f^k\ve{\xi}^k,\ve{s}^k) S^k\right] \ket{\Psi(\Sigma_{k-1}; \Lambda_{k-1})}\nonumber\\
 &=& \left[P(X^k,f^k\ve{\xi}^k,\ve{s}^k) S^k\right] \tilde{T}_k\tilde{T}^\prime_k \ket{\Psi(\Sigma_{k-1}^\prime; \Lambda_{k-1})}\nonumber\\
 &=& \tilde{W}_k\left[P(X^k,\ve{\xi}^k,\ve{s}^k) S^k\right]\tilde{T}^\prime_k \ket{\Psi(\Sigma_{k-1}^\prime; \Lambda_{k-1})}\nonumber\\
 &=& \tilde{W}_k\tilde{W}^\prime_k \left[P(X^k,f^{\prime k}\ve{\xi}^k,\ve{s}^k) S^k\right] \ket{\Psi(\Sigma_{k-1}^\prime; \Lambda_{k-1})}\nonumber\\
 &=& \tilde{W}_k\tilde{W}'_k\tilde{R}_k\tilde{R}'_k \left[P(X^k,f^{\prime k}\ve{\xi}^k,\ve{s}^k) S^k\right]\nonumber\\
 &&\hspace{10em}\ket{\Psi(\Sigma_{k-1}^\prime; \Lambda_{k-1})}\nonumber\\
 &=& \tilde{W}_k\tilde{R}_k\tilde{W}'_k\tilde{R}'_k \ket{\Psi(\Sigma_{k}^\prime; \Lambda_{k})}\nonumber\\
 &=& \tilde{T}_{k+1} \tilde{T}'_{k+1} \ket{\Psi(\Sigma_{k}^\prime; \Lambda_{k})},
\end{eqnarray}
up to a global phase.   This is exactly ({\ref{appendix-base-step}) for $\alpha = k$, and therefore
we reach (\ref{general-graph-main-statement}) by mathematical induction.

\end{document}